\documentclass[doublecol]{epl2} 

\title{Radiation damage in biological material: electronic properties and 
electron impact ionization in urea}
\shorttitle{Radiation damage in biological material} 

\author{C. Caleman\inst{1} \and C. Ortiz\inst{2} \and E. Marklund\inst{3} 
\and~{F. Bultmark}\inst{2} \and M. Gabrysch\inst{4} \and F. G. Parak\inst{1} 
\and~{J. Hajdu}\inst{3} \and M. Klintenberg\inst{2} \and N. T{\^ i}mneanu\footnote{corresponing author, email: nicusor@xray.bmc.uu.se}\inst{3}}
\shortauthor{C. Caleman \etal}

\institute{                    
  \inst{1} Physik Department E17, Technische Universit{\"a}t M{\"u}nchen - James-Franck-Strasse, DE-85748 Garching, Germany\\
  \inst{2} Department of Physics and Material Science, Uppsala University -  {\AA}ngstr{\"o}mlaboratoriet, Box 530, SE-751\,21 Uppsala, Sweden\\
 \inst{3} Department of Cell and Molecular Biology, Uppsala University - Biomedical Centre, Box 596, SE-751\,24 Uppsala, Sweden\\
\inst{4} {Department of Engineering Sciences, Uppsala University - {\AA}ngstr{\"o}mlaboratoriet,
Box 534, SE-751\,21 Uppsala, Sweden}
}

\abstract{
Radiation damage is an unavoidable process when performing structural
investigations of biological macromolecules with X-rays.
In crystallography this process can be limited through damage distribution
in a crystal, while for single molecular imaging it can be outrun
by employing short intense pulses.
Secondary electron generation is crucial during damage formation and we
present a study of urea, as model for biomaterial.
From first principles we calculate the band structure and energy loss function, and
subsequently the inelastic electron cross section in urea.
Using Molecular Dynamics simulations, we quantify the damage and study
the magnitude and spatial extent of the electron cloud
coming from an incident electron, as well as the dependence with initial energy.
}

\begin{document}

\maketitle

\section{Introduction}

Recent advances in the development of X-ray Free Electron Lasers (XFEL)
 offer a tantalizing ability to do photon science 
using short intense X-ray pulses. As a consequence,
a number of theoretical models describing the physics of extreme
X-ray--material interaction have been 
presented~\cite{Neutze2000a,Bergh2004a,Riege2004a,Jurek2004b,Ziaja2006a,Bergh2008a}. The potential
to use the  XFEL to do 3D single bioparticle 
imaging~\cite{Neutze2000a} has enhanced
the efforts to theoretically describe the 
dynamics of a sample being exposed to an XFEL pulse. On
a time scale longer than 5 fs, most of the ionizations in a sample 
exposed to an X-ray pulse will not be due to primary photo ionizations,
but due to inelastic electron scattering. In the aftermath of a single
photo ionization, the photo electron as well as consecutive 
Auger electrons will interact with outer shell electrons in the 
surrounding atoms, leading to an electron cascade (electron shower). 
This process is exploited when detecting photons in photo-multipliers, but leads 
to undesired ionization and structure damage in diffractive imaging.
It is therefore important to know the inelastic scattering properties
and understand the dynamics of the electron generation when 
describing the damage process. The inelastic cross section 
provides information regarding the deposition of energy from an 
energetic electron into the system. 
For free atoms and ions the  electron cross sections
are known, and there are
a few simple empirical expression presented in the literature 
that give good estimation of the free atomic cross sections, 
see for example ref.~\cite{Lennon1988a}.

Structural determination of biological relevant samples, such as
proteins, viruses or cells, are of large interest in many branches
 of life science. In the near future, facilities 
such as the European XFEL source in
Hamburg and the LINAC Coherent Light Source (LCLS) at 
the Stanford 
Linear Accelerator Center, as well as table-top 
XFEL~\cite{Guenter2007a} have promised
to provide short hard X-ray pulses, at wavelengths suitable for atomic 
or near-atomic
structural determination of biological molecules. To describe the secondary
electron production in biomolecular samples, we  present a theoretical 
study of urea, 
in which the cross sections were derived from first principle calculations,
 using a methodology developed and
tested earlier~\cite{timneanu2004a,Ortiz2007a}.
 
\section{Method}
We
perform simulations of the electron cascade in crystalline urea
 (CO(NH$_2$)$_2$) using 
the spatial electron
dynamics program, \textsc{ehole}, that is a part of the  
\textsc{gromacs}~\cite{Spoel2005a} Molecular Dynamics software package. 
To do these simulations we proceed 
according to the 
following work flow: 
(i) calculate the band structure and energy loss function (ELF), 
(ii) calculate the
elastic and inelastic electron cross sections using the energy loss function for the latter,
and finally
 (iii) simulate the 
electron cascade from a single 
electron at a certain energy.
%
%
\subsection{Band structure and energy loss function from
first principle calculations}
\label{sec:elf}
%
%
The method used to calculate the inelastic cross section 
 is based on the ability to calculate the energy loss function
from first principle calculations. 
Performing such calculations
requires a sample with a relatively few atoms in the unit cell. 
We have chosen to
use urea as model for a biological sample due to three major 
reasons. Urea has a well 
known crystalline
structure~\cite{Pryor1970a}.
Urea has an atomic composition of biological character, 
containing light elements such as carbon, nitrogen, oxygen and hydrogen,
 and having a
density close to the density of cellular constituents.
Furthermore, 
the urea crystal
is among the simplest crystalline organic materials known, with 16 
atoms per unit cell in the tetragonal space group $P\overline{4}2_1m$.

The band structure of an idealized urea crystal is determined from 
first principles using structural data found in the 
literature~\cite{Mullen1978a,Ortiz2008a}. We use the \textsc{abinit} first 
principle computer code \cite{Gonze2002a} to generate
 electronic structure results 
within the Local Density Approximation (LDA).
%
 In 
our calculations, we use a 12$\times$12$\times$12 reciprocal 
space grid and a plane wave cutoff value of 680 eV for standard
 LDA pseudo-potentials. 
Additionally, a scissors operator of 1.46 eV is used to correct the 
underestimation of 
the LDA band gap to optical data~\cite{Donaldson1984a}.

\subsection{Electron cross sections}\label{sec:crosssect}
The generation of inelastic electron cross sections
from energy loss functions is well known, and several methods are found in 
the literature.
The method we use here is the so called 
Tanuma-Powell-Penn model (TPP), based on Bethe~\cite{Bethe1930} and 
Lindhard's~\cite{Lindhard1954} work. We refer the reader to 
 references~\cite{Penn1976,Penn1987a,Tanuma1988,Tanuma1993b} 
for a detailed information about the model. {In earlier 
publications we have, in addition to TPP,
employed an alternative model developed by Ashley 
{\em et al.}~\cite{Ashley1988,Ashley1990,Ashley1991}. In the present
 work we  chose to use only the TPP approach, since 
Ashley model was found to underestimate the number of generated 
electrons~\cite{Ziaja2005a}.}
 The approach of using calculated energy loss functions 
when determining the inelastic 
cross sections and 
secondary electron cascades there from, has been 
presented and evaluated in the case 
of two semiconducting materials in an earlier work~\cite{Ortiz2007a}. 
To our knowledge, experimentally determined electron loss functions
for urea are not available  and we are in this work obliged to 
rely on an ELF calculated from first principles for the generation of
the electron cross sections. 
{Estimations of the inelastic electron cross sections
have been presented earlier, both
based on calculated and experimentally derived optical constants for 
biological material~\cite{Laverne1995a}, as well as
direct fitting to inelastic 
experimental data~\cite{Tanuma1993a,Tan2004a,Tan2006a}.
Pioneering work in determining the energy loss function 
in biomaterial from 
first principle calculations was initiated by Zaider 
and coworkers~\cite{Fung1994a}.} 
Tanuma {\it et al.}~\cite{Tanuma1993a} present a study of 
the electron mean free paths between 50 and 2000 eV for
14 organic compounds, comparing and evaluating two fittings to the
so called TPP approach (employed in the present work). 
 
The elastic cross section is calculated
with the Barbieri/van Hove Phase Shift package using a partial
wave expansion technique~\cite{Hove,Bransden1998a}.
The elastic cross section has limited impact on the final
characteristics of the generated cascades. Further details on
these calculations are presented in~\cite{timneanu2004a,Ziaja2001a,Ziaja2005a}.

\subsection{Electron generation and dynamics}\label{sec:md}
Once we have determined the electron cross sections, 
we simulate the electron cascade 
as follows: (i) an initial electron with a certain energy 
is generated. For each time step we calculate the probability for 
inelastic or elastic collisions from the cross sections. The probability 
for any collision event is compared to a random number which determines
whether the electron is scattered or not. (ii) When the electron scatters 
inelastically 
the energy loss of the electron 
is possible through two channels: (a) if the energy loss is higher than 
the binding energy of the outer shell electron
%
%
($E_{\mathrm{lost}}>E_{\mathrm{bind}}$), 
an electron carrying
a kinetic energy equal to the energy lost by the first electron 
minus the binding energy ($E_{\mathrm{kin}}=E_{\mathrm{lost}}-E_{\mathrm{bind}}$)
is released. Now the system contains two electrons with the energy of the 
initial electron minus the binding energy of the second electron. 
These two electrons are now separately treated as 
the initial electron was. The cascades evolve independently 
for each electron.
(b) If the energy loss by the initial electron is less than the 
binding energy ($E_{\mathrm{lost}}<E_{\mathrm{bind}}$), the initial 
electron changes its path and loses the energy to the system.  
(iii) In the case of an 
elastic scattering event, the initial electron keeps its 
kinetic energy and 
changes its path, according to the elastic differential cross section.
The generation of new electrons through impact ionization carries 
on until all electrons have a
kinetic energy lower than the binding energy. 
Each electron 
trajectory is
explicitly followed, and the kinetic energy and momentum
 of each individual electron
is known at every time step. A time step of one attosecond was used 
in the simulations.  


For computational simplicity we have assumed that all 
energy losses in the scattering events are absorbed by the system.
The model does not include hole scattering, electron-hole recombinations, nor
electron-phonon coupling.
Neglecting the scattering of the holes accounts for a lowering 
of the total number of the ionizations events.
%
In diamond this has been shown to underestimate the total number of 
electrons generated nearly by a factor of two compared to 
experiments~\cite{Gabrysch2008a,Ziaja2005a}. 
Previous results for water, however,
have shown good agreement with radiolysis data from experiments,
albeit 
the measurements were performed
on longer time scales~\cite{Muroya2002a}. 

%

Electron--hole recombination -- where electrons and holes recombine
and deposit energy into the system as phonons, photons, or excitations -- is in
principle part of the scattering dynamics. However, the recombination
coefficient for urea was not found in the literature, and since the
coefficients for other organic semiconducors span many orders of
magnitude (see e.g. 
refs.~\cite{Silver1966a,Kalinowski1998a,chen2005c}), estimates from 
analogous compounds
will most likely be gravely misleading. 
We have chosen to
ignore recombinations in our calculations, knowing it could
potentially overestimate the number of ionizations, as electrons
that would recombine cannot generate more electrons. However, the impact
on the results will be limited as long as the carrier density is moderate,
which is the case for all the electron cascades generated in this study.

In an earlier experimental study we have followed the formation of 
secondary electron cascades in single-crystalline diamond caused by
femtosecond X-ray pulses~\cite{Gabrysch2008a}. The experimental results
showed good agreement with a theoretical calculation based 
on the TPP model~\cite{Ziaja2005a}, when excluding 
electron--hole recombination. 
Furthermore, as shown in the next section, the large 
band gap calculated for urea 4.74 eV 
 is close to that of diamond
5.46 eV. Since the density of crystalline urea is lower than that of diamond
($\rho_{\mathrm{diamond}}\approx$~3.5 
g/cm$^3$, $\rho_{\mathrm{urea}}\approx$~1.37 g/cm$^3$), we expect the
electron--hole recombination in urea to be lower than in diamond.

{
Estimations of the vibrational spectra and the electron-phonon coupling strength 
of the urea crystal require an accurate evaluation of the forces acting on the 
molecule upon intermolecular displacements. In its solid phase, urea has a planar 
configuration with vibrational modes ranging between 2.5 meV to 439 meV at 10 K~\cite{Johnson2003}. 
Hence, the energy losses are expected to be somewhat larger than for diamond~\cite{Ziaja2005a}, 
but neglected here.}

{
 Dissociative
Electron Attachment (DEA), {\em i.e.} fragmentation following electron capture 
at subionization energies, has been demonstrated to play an important role in
radiation damage of biomolecular 
samples\cite{Boudaiffa2000a, Huels1998a}.
No experimental DEA cross sections for urea are known, and although cross sections
for several biomolecules can be found in literature, the majority are for the
gas phase. Experimental DEA cross sections for 
tetrahydrofuran~\cite{Breton2004a}
-- a rudimentary analogue to the deoxyribose rings in DNA -- and acetone~\cite{Lepage2000a},
both in the condensed phase, are nearly zero below 6 and 8 eV, respectively,
and have peaks that are almost two orders of magnitude less than our calculated
elastic cross sections in the low energy regime. We have thus 
concluded that omitting 
DEA from our calculations will have no or limited effects on the results.
}

\section{Results and Discussion}
%
%
\subsection{Band structure and energy loss function}
%
\begin{figure}
    \includegraphics[width=8cm]{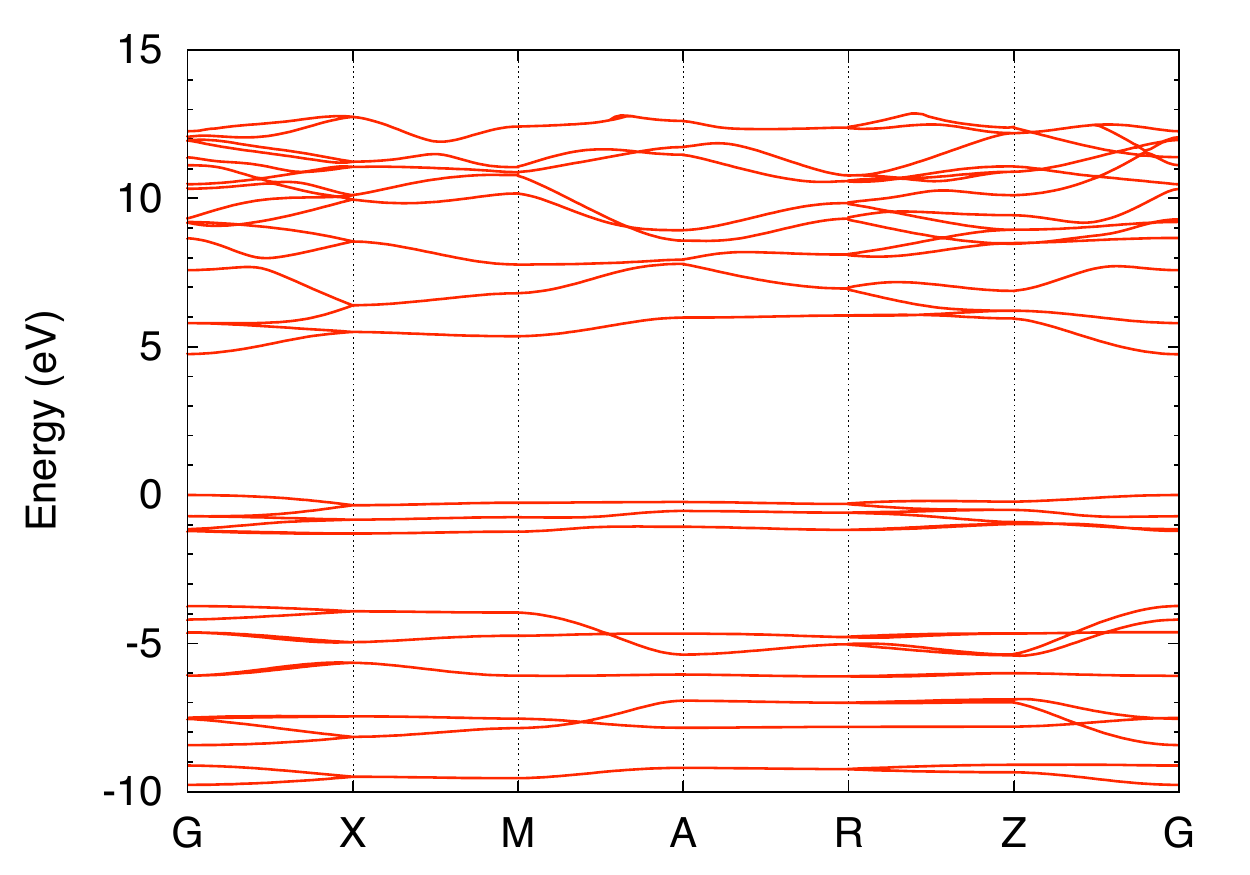}
    \caption{Band structure of urea obtained by Local Density Approximation. The energy bands show small 
dispersion and a direct band gap of 4.74 is observed at the gamma point 
(compared to the experimental value 
6.18 eV~\cite{Donaldson1984a}). A scissors operator of 1.46 eV is used 
to correct
the LDA band gap to experiment and shifts the results to the right. The 
symmetry points in reciprocal space, 
listed 
along the x-axis are G=(000), X=(0$\frac{1}{2}$0),
M=($\frac{1}{2}$$\frac{1}{2}$0),
A=($\frac{1}{2}$$\frac{1}{2}$$\frac{1}{2}$), R=(0$\frac{1}{2}$$\frac{1}{2}$) and
Z=(00$\frac{1}{2}$).}
    \label{fig:spag}
\end{figure}
\begin{figure}
    \includegraphics[width=8cm]{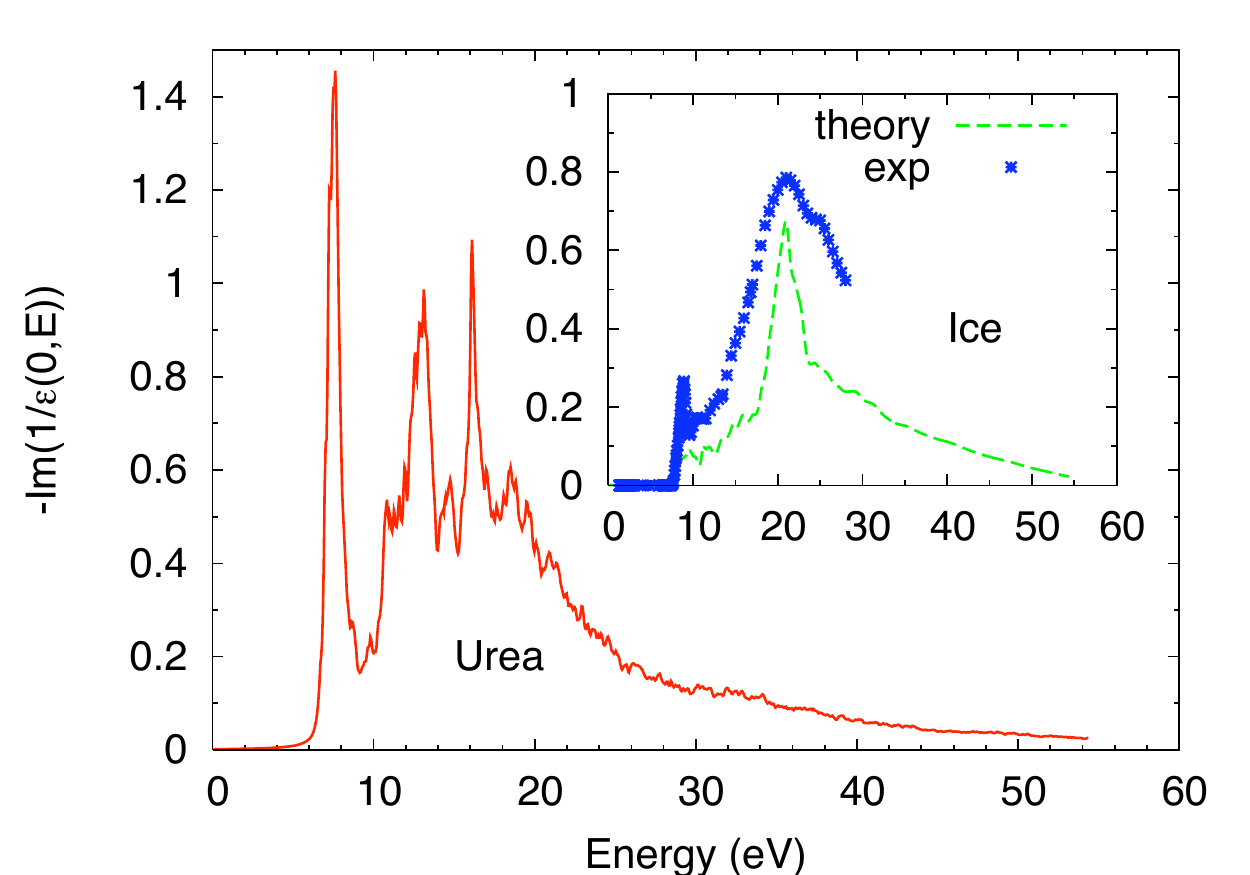}
    \caption{Calculated energy loss function (ELF) of urea. 
 {The inset 
compares the calculated and experimental ELF~\cite{Warren1984} for ice. Due to
the fewer number of van der Waals interactions in crystalline urea as compared to
ice, we expect 
the calculations for urea to show better agreement with 
experiments than ice does.} 
The first plasmon peak in urea is observed at 7.65 eV. }
    \label{fig:elf}
\end{figure}
The obtained LDA  band structure is illustrated 
in fig.~\ref{fig:spag},
by the dispersion 
of eigenstates along high symmetry directions 
for the urea crystal in reciprocal space.
The dispersion is found to be in good agreement with results 
found in the literature.
First principle
calculations on urea have been presented 
earlier~\cite{Dovesi1990a,Lin2003a}. 
There are some
minor disagreements between the band 
structure presented in fig.~\ref{fig:spag}, and those presented in 
ref.~\cite{Dovesi1990a,Lin2003a}, however due to the integration over the
energy loss function with respect to energy 
(see ref.~\cite{Ziaja2001a} for details), 
the impact of small differences in
the band structure on the calculated inelastic
cross section is negligible. 

A direct band gap of 4.74 eV is found at the gamma point 
(6.18 eV experimentally~\cite{Donaldson1984a}), 
other parameters calculated 
are the Fermi energy with respect to the bottom of the valence band ({$E_{\mathrm{f}}$ = 12.81 eV})
 and the valence band 
width ({$E_{\mathrm{v}}$ = 9.71 eV}). 
The energy loss function for zero momentum transfer is to our 
knowledge a feature absent in the literature,
we obtain it from a linear response analysis
within the random phase approximation and it provides a detailed
structure for the energetics of the crystal within low-energy regime.

The calculated energy loss function for urea, 
\mbox{$Im[-\epsilon(q=0,E)^{-1}]$} is presented in 
fig.~\ref{fig:elf}.
{In the absence of experimental 
data for urea, we provide a calculated ELF for ice
and compare it to experimental data (inset in fig.~\ref{fig:elf}).
First principle calculations of molecular crystals, 
where hydrogen bonds and van der Waals interactions are abundant, 
has been much debated in the past years without a consensus being reached. 
The debate regards which parametrization of the exchange and correlation 
contributions to the Coulomb potential is the most suitable, 
see for example reference~\cite{Civalleri2007} for a discussion on cohesive energies. 
Our LDA band gap and the characteristically flat dispersion of eigenstates 
is a good indication that the optical transition probabilities 
are being well described. Also, the close agreement between our calculations 
and experimental data for ice~\cite{Warren1984}, as seen in inset in fig.~\ref{fig:elf},
gives our urea results further support, 
as does a comparison with polystyrene~\cite{Tan2006a} calculated based on a fit to optical data~\cite{Inagaki1977}.
}
For higher energies ($>$50 eV) the cross sections are
generated 
within the Free Electron Gas approximation.
\subsection{Inelastic electron cross section}
\begin{figure}
\includegraphics[width=8cm]{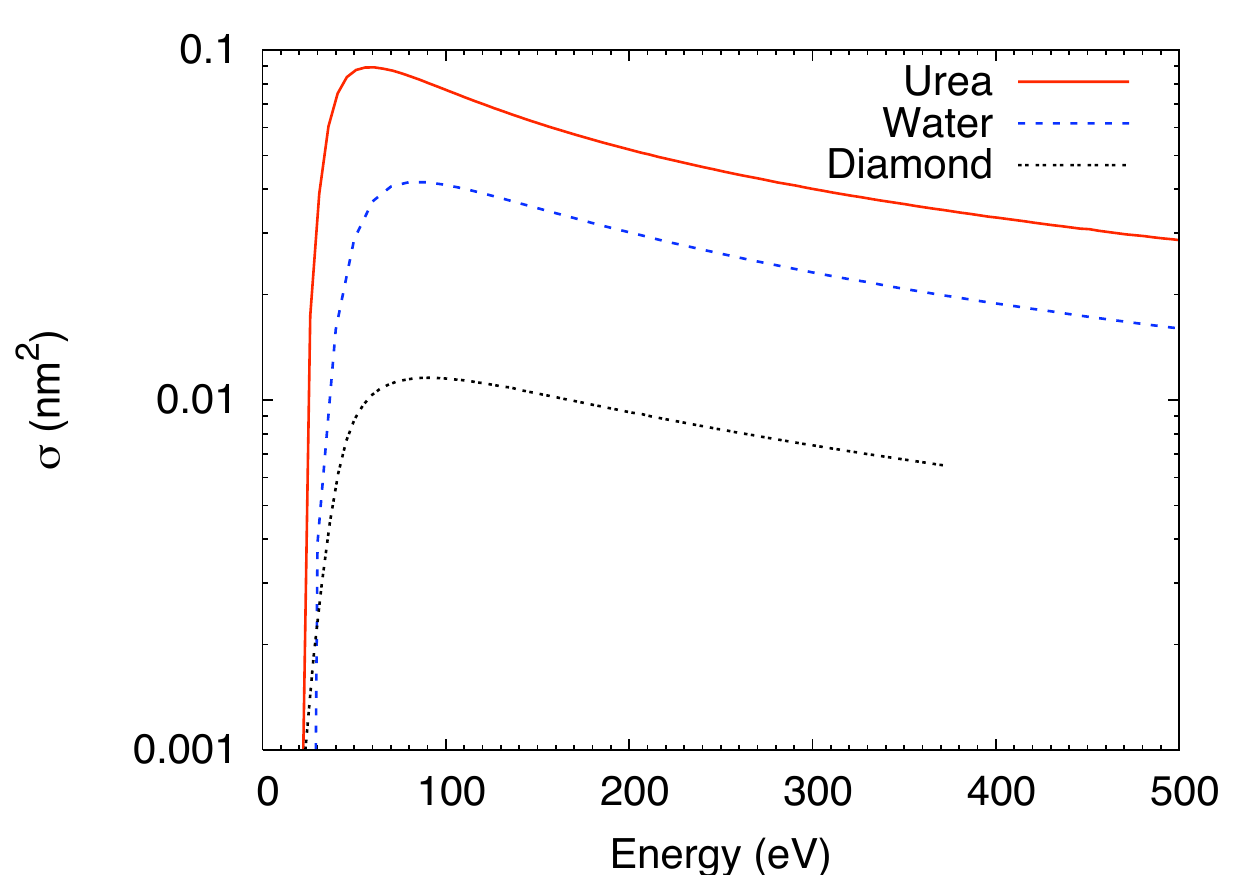}
   \caption{Inelastic cross-section for electron scattering in urea. 
Diamond and water are plotted as a comparison, taken from 
ref.~\cite{timneanu2004a}
and~\cite{Ziaja2001a} respectively.} 
  \label{fig:sigmainel}
\end{figure}

The electron inelastic cross section for urea, together with those of
water and diamond are presented in fig.~\ref{fig:sigmainel}. 
The water and diamond cross sections are presented as a comparison 
(taken from ref.~\cite{timneanu2004a}), both based on experimental
energy loss functions. 
Being slightly denser than water, 
however with similar electronic properties, we expect   
the inelastic electron cross section to be higher than that
for water (fig.~\ref{fig:sigmainel}). 
Comparing the inelastic electron cross section 
for urea to the estimated cross sections for other organic compounds, such as
nucleic acids~\cite{Tan2004a,Tan2006a}, we can also conclude that 
our calculations agree with earlier estimations of similar compounds.
As compared to inelastic cross sections for 
free atoms, the cross sections for diamond (calculated
using the approach employed in this work~\cite{Ziaja2001a}) 
is around 50\% lower than that for free carbon atoms~\cite{Brook1978a} 
(experimentally determined).
%
%
%
\subsection{Ionization and electron dynamics}
\begin{figure}
\includegraphics[width=8cm]{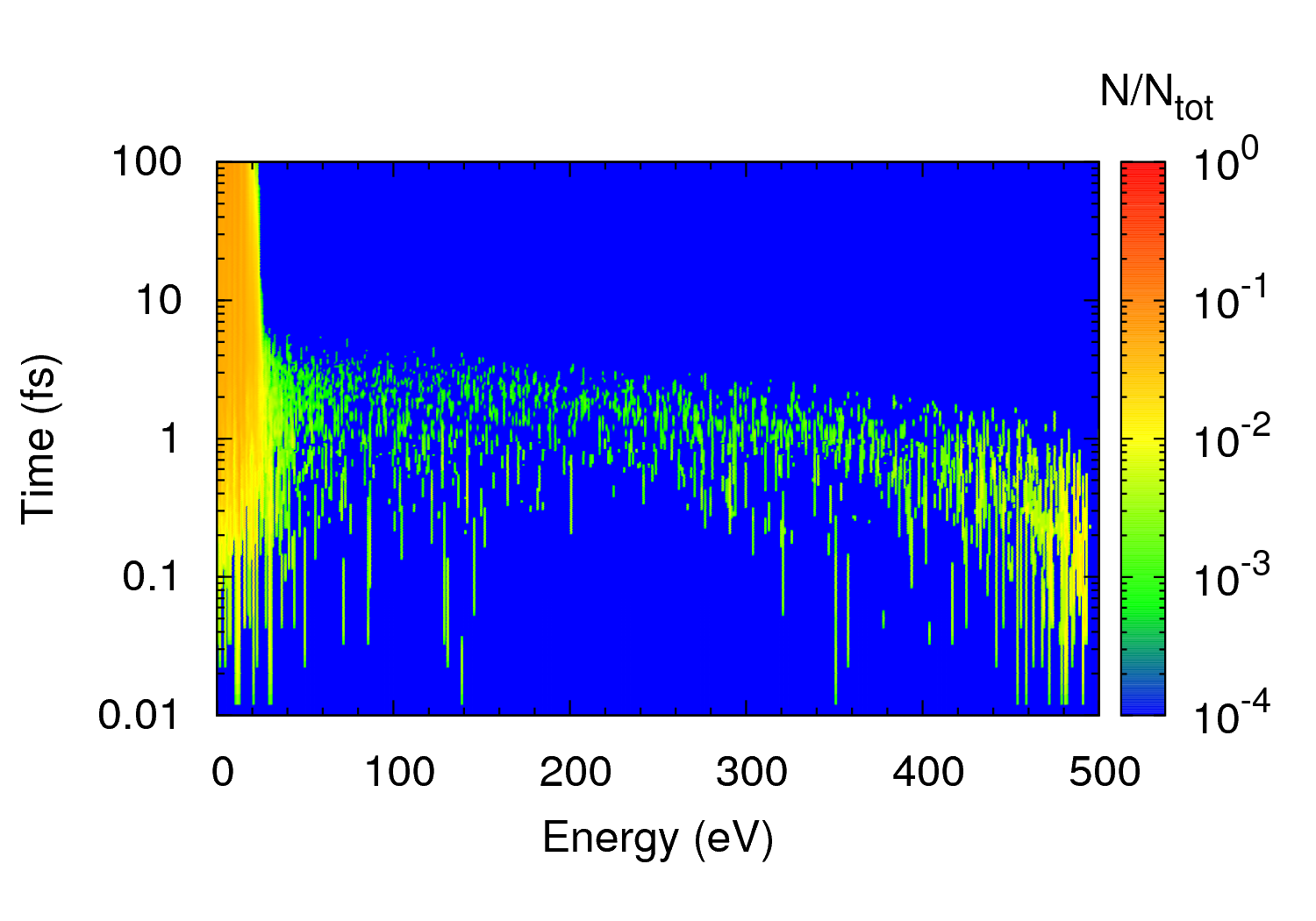}
   \caption{Time evolution of the energy distribution in secondary electron cascade, 
   generated
from an initiating electron with a kinetic energy of 500 eV. The total 
number of 
electrons is normalized to 1 for every timestep (by dividing by the 
total number of electrons in the system).  The figure 
shows an average over 100 simulations.}
  \label{fig:energy_3d}
\end{figure}

%
%
In contrast to earlier calculations of secondary electrons in 
water~\cite{timneanu2004a} and diamond~\cite{Ziaja2001a}, urea generates more 
secondary electrons from an initiating electron
with the same energy. An electron with a kinetic energy of 278 eV generates
slightly above 12 secondary electrons in urea, whereas in water and diamond
only 9 and 6 electrons are generated respectively. {The number of
electrons generated from a single impact electron for a certain
compound is known to be related to the band gap of the 
material~\cite{Klein1968a,Ziaja2005a} as 
$N_{\mathrm{electron}}\sim E_{\mathrm{init}}/(CE_{\mathrm{gap}})$, where $C$
is a compound specific constant. 
}
Fig.~\ref{fig:energy_3d} illustrates how the average energy in the
electron cloud drops down and equilibrates within a few
femtoseconds. The figure shows the evolution of the energy
distribution from an originating electron of 500 eV.
The 
electron cascade in urea {reaches energies below the ionization
threshold} after around 5 fs, which is 
twice as fast as seen in water. {The electron
energy loss}
is related to the density of the crystal, and urea is therefore
expected to equilibrate faster than  water. The main Auger decay lines
in urea are at around 250 eV (carbon), 400 eV (nitrogen) and 500 eV (oxygen). 
For these energies, the electron cascade generated by the Auger electron comprises of
approximately 12 electrons (carbon), 18 electrons (nitrogen) and 24 electrons (oxygen).

In a small biomolecular sample exposed to an X-ray beam, the photoelectron is likely to
escape the system, and it is the Auger electron only that will generate 
the secondary electron cascade. In  such systems the sample will
generate approximately as many electrons as a pure water sample, as water has
slightly lower average Auger energy and a slightly lower density. 
This makes water a good model system for simulating X-ray damage as 
was done by Bergh {\it et al.}~\cite{Bergh2004a}. For larger systems 
however, the 
photoelectron does not escape, 
and using water as a model system  
underestimates the number of generated electrons by about 30\%.  
   
\begin{figure}
    \includegraphics[width=8cm]{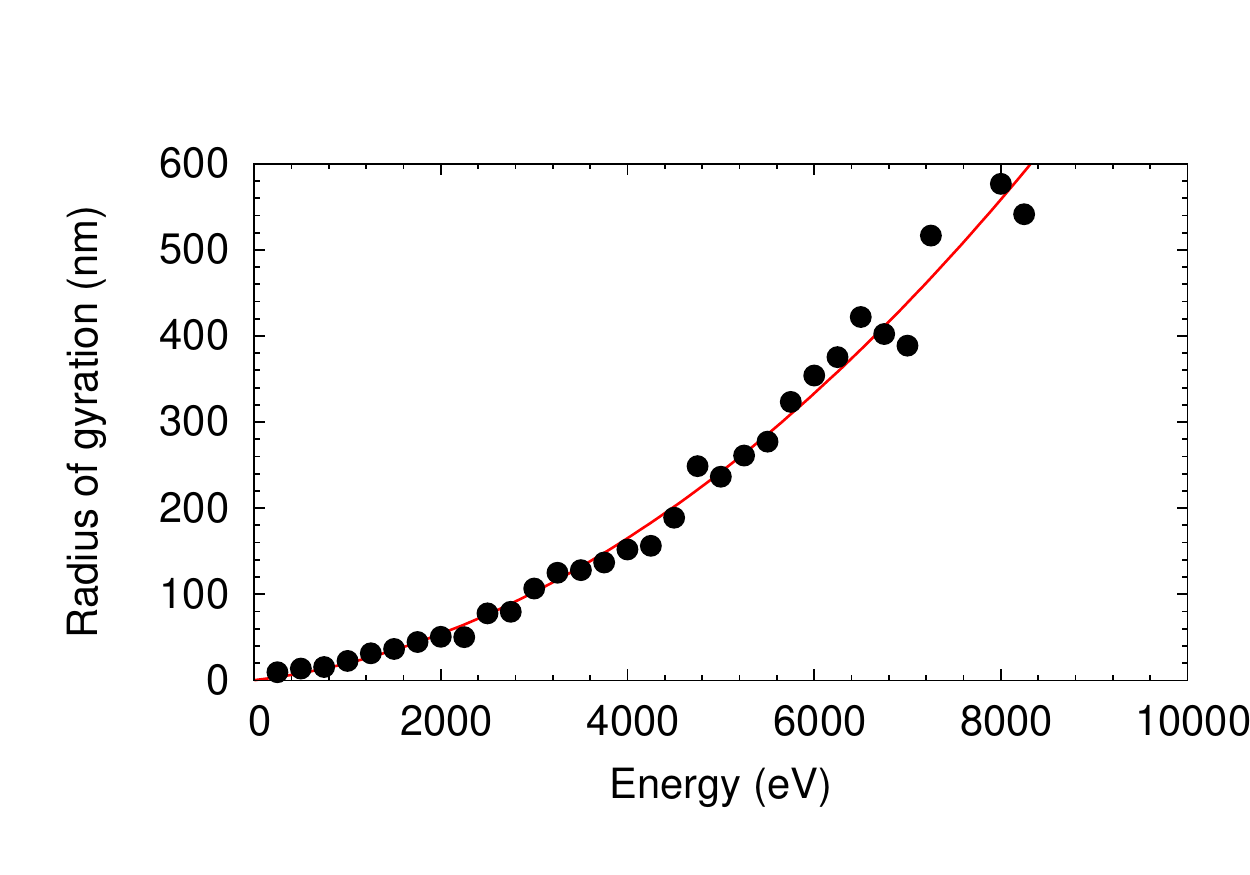}
    \caption{Radius of gyration (average distance from center of mass)
of the electron cloud as a function of the energy of the
initiating electron, after 50 fs. Each dot represents 
the average over 50 simulations,
the line is a fit to the dots, a second order polynomial: $y(x)=0.18+0.013 x + 7.14\times 10^{-6} x^2$.}
    \label{fig:gyrate_all}
\end{figure}

In the energy regime that is investigated in the present work, we
 found that the 
total number of electrons generated after {they
have attenuated to subionization energies}
is linearly 
dependent on the energy of the initiating electron, as follows
\begin{equation}
N_{\mathrm{electron}}= \frac{E_{\mathrm{init}}}{4.5E_{\mathrm{gap}}}.
\end{equation} 
Similar behaviour has been 
shown before {for 
semiconductors~\cite{Ortiz2007a,Klein1968a,Alig1980a} 
as well as for diamond~\cite{Ziaja2005a}.}
  Damage due to
ionization is related to the density of ionization events. The radius 
of gyration
{({\it i.e.} 
the average distance of the electrons to the center of mass)} 
of the generated electron cloud is not linear dependent on the 
initiating electron energy, but rather as a second order
polynomial of the 
energy (fig.~\ref{fig:gyrate_all}). This implies that the electron cloud from
a low energy electron is much more dense, than that from a high energy 
electron, and therefore more damaging, despite the lower number of 
total electrons generated. {The size of the ionizing
electron cloud is quantified by the radius of gyration of the cloud.  
This relates to the energy of 
the initiating electron, which serves as a basis for understanding
how the damage process depends with size in various crystals.} 

Since the cross sections are calculated for a neutral system,
we do not explicitly create positive charges in the system and 
only account for the number of electrons lost. 
{For a highly ionized system, the use of atomic cross sections will no 
longer be accurate, a plasma approximation is then needed~\cite{Bergh2008a}.} 
This lapse in the model is likely to have 
an impact on the total number of electrons generated and remains to be studied. 
Furthermore, omitting the positive charged ions will lead to
a larger electron cloud than expected, since 
we cannot account for shielding effects.
However, for a system that is only fragmentary ionized,
the impact of disregarding the positive charges 
should be negligible. 

\section{Conclusion}
%
We have calculated the band structure and energy loss function for urea crystal
from first principles. Employing a methodology presented and tested earlier,
we have determined the inelastic electron cross section and estimated the 
number of secondary electrons generated from an incident electron at different 
energies.  
{We are confident that the similarity of the ELF 
with experimental data and 
previous calculations on similar
molecules implies our predictions are good models of reality.}
Our simulations show that
urea generates 30\% more electrons than water, from an 
initiating electron carrying the same energy. Such a difference is expected considering the density difference
of urea and water. Urea bears more similarities with the biological matter 
(such as cells, viruses) in both elemental composition and density, making it a more suitable object of
theoretical study than water or diamond, especially when addressing large systems.
Furthermore we describe the size of the electron cloud generated 
from a single electron as a function of the electron energy, showing the 
spatial extent of the radiation damage due to secondary ionization.
Secondary electron generation is an important effect for the damage
 caused on a biomolecular sample by an intense 
X-ray beam, and it is therefore necessary to understand these processes  
in order to predict the damage and aid the structural determination of single molecules or crystals.
\acknowledgements
The following organizations are acknowledged for their financial support: 
the Swedish Research Foundation (thanked by CO, EM, MG, FB, MK, JH, NT), G\"oran Gustafssons Stiftelse (thanked by CO and MK) and the 
 {DFG Cluster of Excellence: Munich-Centre for Advanced Photonics}
(thanked by CC, FGP and JH). 
Although not as cool as ice, or as shiny as diamond, the present work on 
urea has been
endorsed by the following people, to whom we give 
thanks: Magnus Bergh, David van der
Spoel, Richard London and Beata Ziaja. 



\bibliographystyle{../texutil/eplbib}
\bibliography{../texutil/monster}

\end{document}